\documentclass[journal]{IEEEtran}
\IEEEoverridecommandlockouts

\usepackage{amssymb}
\usepackage{float}
\usepackage{makeidx}
\usepackage{amsmath,amsfonts}
\usepackage[tight,footnotesize]{subfigure}
\usepackage{textcomp}
\usepackage{bbm}
\usepackage{epsfig}
\usepackage{epsf}
\usepackage{psfrag}
\usepackage{verbatim}
\usepackage{color}
\usepackage{multirow}
\usepackage{tabularx}
\usepackage{booktabs}
\usepackage{array}
\usepackage{soul}
\usepackage{footnote}
\usepackage{cite}
\usepackage{dblfloatfix}

\usepackage{xcolor}
\usepackage{color, colortbl}
\usepackage[colorlinks,bookmarksnumbered,citecolor=orange,urlcolor=orange]{hyperref}
\usepackage{graphicx}
\graphicspath{{Figures}}
\DeclareGraphicsExtensions{.pdf,.png}
\usepackage{wrapfig}
\usepackage{bigstrut}
\usepackage[english]{babel}

\usepackage{csquotes}

\usepackage{algorithm, setspace}
\usepackage{amsthm}
\usepackage{algpseudocode}
\usepackage{url}
\usepackage{multirow}
\usepackage{float}
 \hypersetup{
     colorlinks=true,
     linkcolor=orange,
     filecolor=orange,
     citecolor=orange,      
     urlcolor=orange,
     }
\usepackage[T1]{fontenc}
\usepackage{balance}

\newcommand{\p}[1]{\medskip \noindent \textbf{{#1}.}}
\newcommand{\eq}[1]{Equation~(\ref{eq:#1})}
\newcommand{\fig}[1]{Figure~\ref{fig:#1}}

\DeclareMathOperator*{\argmax}{arg\,max}

\title{\LARGE

A Modular Haptic Display with Reconfigurable Signals \\
for Personalized Information Transfer
}

\author{
Antonio Alvarez Valdivia$^{1}$, Benjamin A. Christie$^{2}$, Dylan P. Losey$^{2}$, and Laura H. Blumenschein$^{1}$
\thanks{This work involved human subjects or animals in its research. Approval of
all ethical and experimental procedures and protocols was granted by 
Purdue University IRB, Application No. 2025-19.}
\thanks{This work is supported in part by NSF Grants $\#2129201$ and $\#2129155$ and by the NSF Graduate Research Fellowship Program (DGE-1842166).}
\thanks{$^{1}$Mechanical Engineering, Purdue University, West Lafayette, IN 47901. {\texttt{\{alvar168, lhblumen\}@purdue.edu}}}%
\thanks{$^{2}$Mechanical Engineering, Virginia Tech, Blacksburg, VA 24061.
{\texttt{\{benc00, losey\}@vt.edu}}}
}

\begin{document}
\maketitle


\begin{abstract}
We present a customizable soft haptic system that integrates modular hardware with an information-theoretic algorithm to personalize feedback for different users and tasks. Our platform features modular, multi-degree-of-freedom pneumatic displays, where different signal types --- such as pressure, frequency, and contact area --- can be activated or combined using fluidic logic circuits. These circuits simplify control by reducing reliance on specialized electronics and enabling coordinated actuation of multiple haptic elements through a compact set of inputs. Our approach allows rapid reconfiguration of haptic signal rendering through hardware-level logic switching, without rewriting code. Personalization of the haptic interface is achieved through the combination of modular hardware and software-driven signal selection. To determine which display configurations will be most effective, we model haptic communication as a signal transmission problem, where an agent must convey latent information to the user. We formulate the optimization problem to identify the haptic hardware setup that maximizes the information transfer between the intended message and the user’s interpretation, accounting for individual differences in sensitivity, preferences, and perceptual salience. We evaluate this framework through user studies where participants interact with reconfigurable displays under different signal combinations. Our findings support the role of modularity and personalization in creating multimodal haptic interfaces and advance the development of reconfigurable systems that adapt with users in dynamic human-machine interaction contexts.
\end{abstract}




\section{Introduction} \label{sec:intro}

Designing haptic interfaces that are expressive and interpretable for different human users remains a key challenge in human-computer and human-robot interaction \cite{maclean2008foundations}. Many existing haptic systems rely on fixed feedback modalities or predefined signal sets, limiting their ability to accommodate different users, tasks, or environments \cite{seifi2017exploiting, habibian2025survey,roy2024towards}. Recent work has shown that even modest changes in signal encoding --- such as directional resolution or complexity --- can significantly influence how users perceive, interpret, and act upon haptic feedback\cite{valdivia2023perception, tan2020methodology, wang2019multimodal}. These findings emphasize the significance of personalization: the haptic signals that make the most sense for a given user and task may be confusing and inadequate for a different user or changed task. 

\begin{figure*}[t]
\includegraphics[width=1.8\columnwidth]{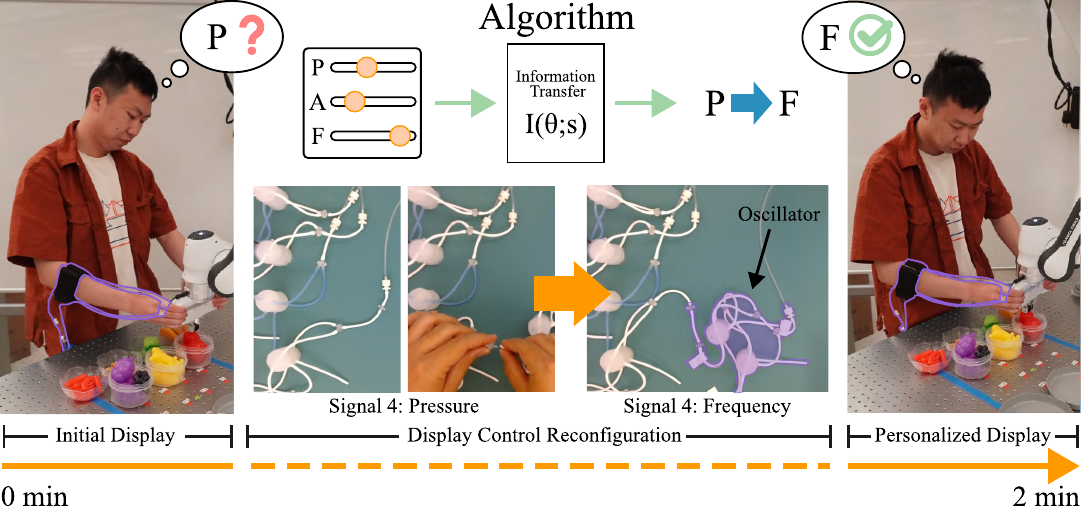}
\centering
\caption{Modular haptic display worn by a human user. The inflation pattern of the haptic display tells the user what the goal is (i.e., which objects to grasp and where to place them). A user may be initially confused about a specific signal type, i.e. pressure. Our proposed approach provides a modular hardware approach supported by an information-theoretic algorithm, so the user can identify a better signal modality, i.e. frequency, and quickly reconfigure the hardware to change the signal type by adding an oscillator.}
\vspace{-1.0em}
\label{fig:front}
\end{figure*}

Consider the setting shown in \fig{front}.
Here a haptic display is trying to communicate the instructions for a recipe, with different signals conveying what ingredient type to grasp and where to add it.
Existing haptic systems rely on a fixed display; e.g., the object to grasp is indicated by vibrations, and the correct location to place it is conveyed by pressure.
This mapping may work well for several users --- but falls short when faced with a wearer who does not accurately discern the differences in pressure (or notice the vibrations).
More generally, users may have different needs and preferences based on experience with the task or sensory sensitivity to the signals \cite{peters2009diminutive, stevens1996spatial}.
Together, these needs have fueled a growing interest in modular, reconfigurable feedback interfaces that support task-driven and user-specific customization \cite{shtarbanov2023sleeveio, pezent2022design, huang2025aerohaptix, huang2023skin}.


To address these challenges, we present a customizable soft haptic system that supports hardware and signal reconfiguration driven by algorithmic personalization. Our \textit{modular hardware} approach combines multi-degree-of-freedom (multi-DoF) soft haptic displays with fluidic logic circuits, which use air flow and passive elements to perform logic operations (see Figure~\ref{fig:front}). Fluidic logic enables sequential or parallel actuation patterns to be encoded directly into the pneumatic hardware, reducing the reliance on electropneumatics to control the system. While fluidic logic has been increasingly adopted in soft robotics to produce complex sequential motions, such as repeated walking patterns \cite{rothemund2018soft,preston2019digital,preston2019soft,kendre2022soft}, its application to haptics remains largely unexplored. We extend the use of fluidic logic to the domain of haptics by showing how its scalability, low power requirements, and mechanical robustness can be leveraged to control and reconfigure multi-channel haptic systems. 
In practice, users can manually switch the logic circuits to reach haptic signal sets with varying levels of pressure, frequency, and contact area (Fig.~\ref{fig:front}). 
This provides a mechanically intelligent platform capable of rendering a diverse library of haptic cues with minimal hardware overhead.

While our hardware enables physical flexibility in signal rendering, selecting which signals to present --- particularly in multi-DoF systems --- requires a systematic evaluation of both informativeness and interpretability. Our modular haptic device enables a wide range of displays: but which ones will be efficient for the current user and desired task? To ensure that signals are interpretable and useful, the system must be \textit{personalized} --- adapting which signals are presented and how they are rendered to align with individual perceptual and cognitive characteristics. To frame this tuning problem, we adopt an information-theoretic perspective \cite{reed1998note, tan2020methodology},  modeling haptic interaction as a communication channel constrained by human perceptual limitations. This information-theoretic framing has also been applied to haptic interfaces, offering tools to reason about how perceptual thresholds, signal complexity, and cue distinctness affect the rate and clarity of communication \cite{tan2009optimum, tan2020methodology, sims2016rate}. These insights suggest that maximizing performance is not simply a matter of increasing signal richness, but of selecting cues that are both salient and easily interpretable. 

Combining both the modular hardware with information-theoretic software, we ultimately present an approach for personalized haptic communication through two complementary layers of modularity.
The mechanical layer enables the rendering of varied signal combinations through reconfigurable hardware; the algorithmic layer determines which of these combinations best match an individual user's perceptual abilities, preferences, and task demands. This integrated framework supports adaptive interactions by maximizing the flow of task-relevant information from system to user.
Importantly, the integrated process accounts for individual differences in perceptual salience, signal sensitivity, and user preference, enabling real-time personalization of the interface.

This work contributes toward the broader goal of making ubiquitous haptic systems that can be applied across multiple users and tasks \cite{gallina2015progressive, beckerle2022embodiment}. 
In doing so, we aim to enable richer, more efficient interactions between users and machines. Overall, we make the following contributions:

\p{Mechanical Customization} We develop pneumatic logic circuits that can render a variety of multi-DoF haptic signals.
By disconnecting and reconnecting different tubes, users can manually adjust to the types of haptic signals rendered by the modular display.
This approach inherently reduces complexity and enables rapid reconfiguration by minimizing reliance on electropneumatic controllers.

\p{Algorithmic Personalization} Our modular hardware sets the stage for a variety of deployments. 
To determine which configuration is best suited for the current user and task, we formulate a signal selection framework grounded in information theory.
Users input their personal preferences and the task specifications, and the algorithm recommends a hardware configuration that should maximize information transfer.
This approach enables the system to personalize the selection of display configurations based on user preferences, perceptual salience, and sensitivity. 

\p{Task-Informed and User-Specific Reconfiguration} We conduct user studies that validate our personalization framework, integrating hardware and algorithmic customization to identify optimal display configurations for specific users and tasks. 
Results across $n=13$ participants and two experimental settings suggest that our modular systems convey information in ways that cause humans to perform tasks more effectively.
Specifically, when we physically deploy the haptic display recommended by our algorithm, users complete the tasks with less error, and subjectively perceive the haptic feedback to be more helpful.

\p{Analyzing Hardware and Software Contributions} 
We perform follow-up tests to understand how each aspect of our modular approach contributes to haptic success.
In terms of the hardware, we find that different users perform best with different configurations; i.e., physically modifying the haptic display is necessary to improve performance.
In terms of the software, we observe that the type of haptic display that users prefer is not always the same haptic display that our algorithm recommends; i.e., maximizing information transfer is not as simple as just choosing signals that people like.
These results highlight that both mechanical modularity and algorithmic personalization are needed to maximize the potential of customizable haptic devices.
\section{Related Works} \label{sec:related}

Our development of modular haptic interfaces connects research on haptics, pneumatic logic architectures, and information transfer.
Below we summarize our intersection with each of these areas.

\p{Modular and Adaptive Haptic Interfaces}
Personalization has become a key design principle in human-computer interaction, with growing recognition that ``one-size-fits-all'' interfaces fail to accommodate differences in user preferences, capabilities, and strategies. Personalization in graphical user interfaces is relatively mature, providing a proof of concept that adapting layout, content, and controls can improve user performance and reduce cognitive load \cite{todi2021adapting, zhao2023towards}. Comparatively, in haptic interfaces, \textit{user} personalization has been narrowly focused on tuning specific signal parameters such as intensity, duration, and modality to better match user needs and perceptual thresholds \cite{choi2022design, yoon2017customizing}.
To adapt instead to changing\textit{ tasks}, prior work has demonstrated modular wearable displays composed of soft composite materials \cite{raitor2017wrap, do2021macro}, patch-based layouts for skin-conformal stimulation \cite{han2023parametric}, room-scale vibrotactile systems designed for spatial reconfiguration and uniform feedback \cite{tsujita2025haptoroom}, and systems that permit geometric or tactile customization through interchangeable elements \cite{spiers2016design, spiers2022s}. These designs emphasize the practicality and versatility of modular architectures, enabling reuse of core components across tasks. 
However, we emphasize that the adaptability of these systems is often limited, allowing a range of designs but not on the fly reconfiguration: for example, actuator modules that can be freely combined into designs but not easily interchanged \cite{shtarbanov2023sleeveio} or actuator attachments that allow arbitrary placement for optimized signal discrimination \cite{huang2025aerohaptix}.
Overall, most existing systems rely on static mappings or handcrafted designs; few integrate reconfigurability, let alone algorithmic personalization \cite{christie2024limit, huang2025aerohaptix} --- limiting their ability to adapt to evolving behavior or task demands.

\p{Fluidic Logic in Soft Robotics and Haptic Systems}
Outside of haptics, fluidic (or pneumatic) logic has emerged as a powerful control paradigm for soft robots and shape-changing systems, enabling fully or partially electronics-free architectures that are robust, lightweight, and mechanically programmable. Foundational work has shown how pneumatic logic gates can be composed into fluidic circuits for signal processing, sequencing, and timing \cite{preston2019digital, preston2019soft, rothemund2018soft}. These systems enable the autonomous control of soft actuators, often relying solely on pneumatic sources and passive structures to execute prescribed behaviors \cite{drotman2021electronics, rajappan2022logic}. Recent efforts have advanced the design of programmable soft valves \cite{decker2022programmable}, bistable elements, and soft ring oscillators \cite{preston2019soft} to support dynamic control tasks without embedded microcontrollers. Sheet-based fluidic diodes and pneumatic code blocks further expand the design space for compact, programmable circuits of modular elements~\cite{vo2024sheet, picella2024pneumatic}.
These innovations have been applied across domains, from robot locomotion \cite{zhai2025monolithic} to smart clothing \cite{rajappan2022logic} to fluidic computing kits for shape-changing displays \cite{lu2023fluidic}. 
Despite this progress, the application of fluidic logic in haptic systems remains relatively underexplored \cite{stanley2024high, jumet2023fluidically}. 

Our work leverages these advances to enable reconfiguration and scalable control of haptic signal outputs, bridging the strengths of pneumatic logic with the challenges of personalized interface design. In creating our system, we prioritized accessibility and ease of replication --- particularly for prototyping environments without access to specialized fabrication tools. Towards this end, we selected a bistable soft valve architecture most related to Preston et al. \cite{preston2019soft}, which has been validated extensively and can be fabricated using off-the-shelf materials. By extending this design into haptic applications we develop novel, modular haptic displays that can be rapidly reconfigured and scaled; all without relying on embedded microcontrollers or advanced manufacturing processes.

\p{Information-Theoretic Approaches to Signal Design} 
Given some physical interface, information theory offers a formal framework for evaluating how that system transfers data to the user. 
Prior works which apply information theory to haptics have realized that effective communication has a human component: human perceptual limitations (e.g., cognitive overload) often constrain information transfer \cite{tan2009optimum, tan2020methodology, spiers2022s, schaack2019haptic}. 
Indeed, variations in signal complexity --- such as modulation of frequency, actuator density, or multi-dimensional encoding --- affect perceptual discriminability and user performance \cite{de2017designing, stock2020feedi, park2018haptic, huang2025aerohaptix}. 
Accordingly, we cannot just calculate the abilities of a haptic display in isolation; its performance inherently depends on the current user and task \cite{sims2016rate}.
Recent approaches have therefore shown that information theoretic metrics (such as information gain) can be leveraged to account for the human's perspective.
This includes selecting questions that are easy for the human to answer \cite{biyik2022learning, habibian2022here}, or determining which types of signal mappings are best suited for the user \cite{christie2024limit, habibian2025survey}.
We emphasize that finding the correct haptic display is not a stationary problem. 
Because different users often interpret the same signals differently, a given haptic display can show inconsistent performance across individuals \cite{belpaeme2018social, gasteiger2023factors}.
In our work, we accordingly apply information theory to quickly determine the appropriate haptic configuration for the current user by accounting for that specific user's preferences and perceptual salience. 

\section{Haptic Device Design and Control}\label{sec:device}

\subsection{The Haptic Displays}

\begin{figure}[t]
    \includegraphics[width=1.0\columnwidth]{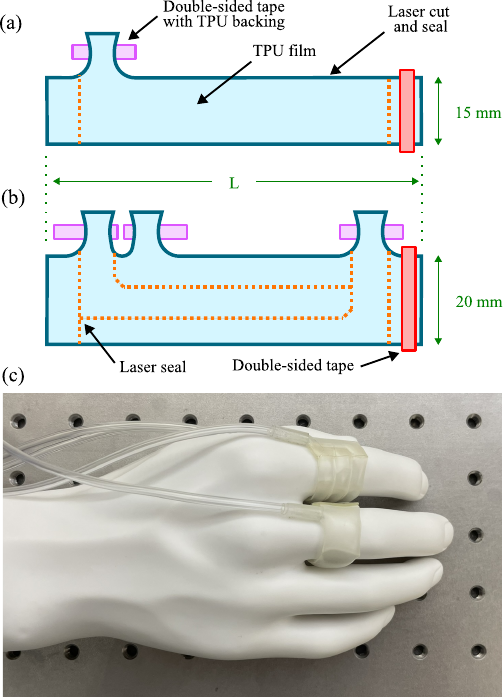}
    \centering
    \caption{The haptic displays are made of layered TPU film, laser cut and sealed following the patterns shown in (a) for the single chamber and (b) multi-chamber versions. In (c), we show the two display models actuated.}
    \label{fig:displays}
    \vspace{-0.5em}
\end{figure}

The haptic displays presented in this work are soft, wearable actuators inspired by previous soft wrapped haptic displays \cite{valdivia2023perception, valdivia2023wrapping, jumet2023fluidically, agharese2018hapwrap, raitor2017wrap}. Designed as ring-shaped interfaces worn on fingers, these devices provide tactile feedback through pneumatic actuation, delivering compression-based haptic sensations similar to a squeeze \cite{endow2021compressables}. 

The displays are fabricated using two layers of thin, heat-sealable thermoplastic polyurethane (TPU) film (HM65 $0.15$~mm $78A 40"$, Perfectex). The elasticity of the TPU film used in these displays provides pronounced tactile sensations under inflation pressures, producing more effective feedback compared to displays made of non-stretchable films. Unlike earlier iterations of soft wrapped haptic displays fabricated with a linear heat sealer \cite{valdivia2023wrapping}, these displays are patterned, cut, and sealed using a $100$~W $\text{CO}_2$ laser cutter (Epilog Fusion Pro 36), allowing for high precision fabrication. 

We begin the fabrication process by cleaning the TPU sheets with isopropyl alcohol (IPA), layering them one on top of the other, and laminating them together with a heated roller laminator. This preparation process will temporarily adhere the TPU layers so they will remain attached during the sealing and cutting process. An acrylic plate with double-sided masking tape is then used as a substrate to hold the laminated TPU films and the plate is placed in the laser cutter. Depending on the speed and power of the laser cutter as it traces out the patterns it will either permanently seal the layers together or both seal and cut. Laser settings for \textit{sealing} were $100\%$ speed, $9\%$ power, and $100\%$ frequency, and for \textit{cutting} they were $75\%$ speed, $17\%$ power, $50\%$ frequency. Figure~\ref{fig:displays} shows the flat patterns used for each display with specification of which lines were sealed and cut. After laser processing, the sealed shape is removed from the substrate. The sealed inlet -— designed with a trumpet-like geometry to facilitate tube insertion -— is carefully opened with a blade to allow insertion of clear, soft PVC tubing ($1/8"$ OD, Masterkleer). The tubing is secured in place using a strip of double-sided viscoelastic adhesive tape (MD-9000, Marker Tape) with one side laminated to a TPU film (to create a single-sided tape). Finally, to assemble the display into its ring shape, the end tabs of the cut structure are joined using plain double-sided viscoelastic tape, forming a closed loop that wraps around the finger.

We developed two configurations: a 1-DoF display composed of a single pouch (Figure~\ref{fig:displays}(a)), and a 3-DoF display with three independently actuated pouches (Figure~\ref{fig:displays}(b)). The multi-DoF version was designed to explore more complex tactile cues, specifically the contact area-based modality described in later sections. The displays were fabricated in varying circumferences ($L = 65$–$90$~mm) to accommodate different finger sizes. These displays can each accommodate approximately $3.75$~mL of internal volume when inflated to $27.58$~kPa, based on geometric measurements taken at full inflation, assuming an elliptic torus with semi-axes of $5$~mm and $12.72$~mm. All displays safely operate at pressures up to $34.47$~kPa.

\subsection{Fluidic Logic Control}
\label{subsec:fluidic_logic}

\begin{figure*}[t]
    \includegraphics[width=2.0\columnwidth]{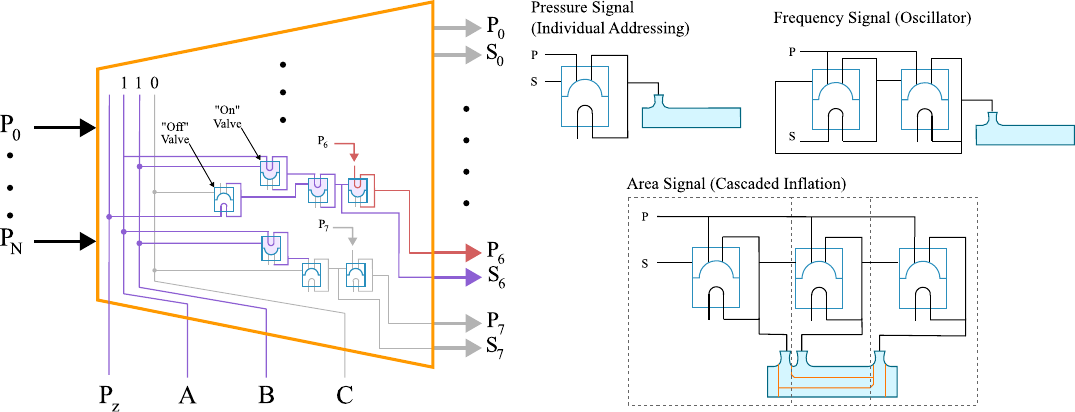}
    \centering
    \vspace{-0.5em}
    \caption{Fluidic logic control architecture for reconfigurable haptic displays. The system is divided into two stages: code logic (left) and output logic (right). The code logic functions as a pneumatic demultiplexer, converting a three-bit input (A, B, C) and a constant logic-high pressure source $P_Z$ into up to eight distinct output states via soft valves operating as logic gates, enabling selective pressure routing of an actuation source $P_i$. The output logic stage implements three actuation modes: (1) \textbf{pressure signals} via direct inflation of a single output, (2) \textbf{frequency signals} using a soft oscillator composed of valve pairs, and (3) \textbf{area signals} achieved through cascaded inflation of multiple chambers. For the area signal, the dotted boxes indicate that additional stages of valves and displays can be added to extend the cascade, allowing for larger or more spatially distributed tactile patterns.}
    \label{fig:fluidic_logic}
    \vspace{-0.5em}
\end{figure*}

\subsubsection{Soft Valve Design and Characterization}
The haptic displays are controlled using soft fluidic logic circuits made up of elastomeric valves with two states, inspired by the soft valves demonstrated by Preston et al. in their soft ring oscillator \cite{preston2019soft}. Although originally implemented as pneumatic inverting Schmitt triggers, here we exploit their general bistable behavior -- switching reliably between two distinct states using pneumatic inputs -- to control airflow pathways. 
After surveying the range of design for fluidic logic in literature, these soft valves stood out due to their widespread validation, accessible fabrication process, use of readily available materials, detailed supplementary documentation that facilitated replication, and compatibility with the working pressure range of our soft haptic displays. We made slight modifications to the original valve design to simplify the fabrication process while maintaining functional performance. 

The original STL files for the mold of the soft valves used for the inverter \cite{preston2019soft} were obtained from the supplementary materials of that paper. In the original implementation, the internal channels were molded. Instead, we modified the valve to use stock silicone tubing ($1/16"$ ID X $1/8"$ OD, shore hardness $35A$, McMaster-Carr) for the internal channels by expanding the junction between the tubes to accommodate for the larger tubing diameter and thickening the top and bottom flat face molds to account for the increased stiffness. The fabrication process for assembling the internal tubing followed the approach in \cite{preston2019soft}. Smooth-On: Dragon Skin 10 NV (Smooth-On) was used for molding the flat faces and cylindrical wall-membrane assembly, and Smooth-Sil 950 (Smooth-On) was used for the junction and end cap of the internal tubing. For bonding the parts together, we used silicone epoxy (Silpoxy, Smooth-On), which cures faster than uncured elastomer.

Valve characterization confirmed consistent snap-through pressures, actuation speeds, and airflow capabilities. Using a pressure sensor (015PGAA5, Honeywell Sensing) connected to the inlet of the upper chamber of the valve, the mean snap-through pressure was measured at $11.44\pm2.29$~kPa across 5 different valves. To ensure robust performance throughout the system and to compensate for tubing losses, all logic operations in the code logic stage were conducted at $20.68$~kPa, well above the snap-through threshold, turned on and off using solenoids (ET-3-6, Clippard). The average airflow through the open valve channels was $2.76\pm0.28$~slm (standard liters per minute), measured after snap-through across 5 different valves using an airflow sensor (PFLOW3008, Angst+Pfister). The rise time of this flow was approximately 0.51~seconds from initial pressurization measured to a downstream valve.

\subsubsection{Fluidic Control Architecture}
Our fluidic control architecture consists of two primary stages: the \textit{code logic} and the \textit{output logic}. The \textit{code logic} functions as a pneumatic demultiplexer, translating a three-bit input into eight possible output states, allowing control of up to seven discrete signals with one reserved state for no actuation. Logic gates (NOT, AND, OR) are constructed from these soft valves \cite{preston2019digital, kendre2022soft}, processing binary pressure inputs (A, B, C) and a constant logic-high pressure source ($P_z$) to selectively route actuation pressure ($P_0, P_{1}, ..., P_{N}$) provided by hand pressure regulators (6763K82, McMaster-Carr) to the \textit{output logic} stage. Additional details of the logic designs used to accomplish the demultiplexing are included in Appendix~\ref{appendix}. Depending on the amount of individual signals that are wanted, the \textit{code logic} can be adjusted down to a $2$ input - $4$ output setup, or up to a $4$ input - $16$ output system. 

The \textit{output logic} stage provides three different actuation methods for specific tactile feedback scenarios: (1) \textit{individual addressing}, using single valves to independently turn on or off airflow from pressure sources; (2) \textit{oscillator}, combining valves to create oscillating inflation with pressure dependent frequency; and (3) \textit{cascaded inflation}, employing sequential valve activation to inflate multiple chambers in a controlled order. These diverse configurations, shown in Figure~\ref{fig:fluidic_logic}, enhance the versatility of the haptic feedback qualities delivered by our reconfigurable, modular system. This architecture enables rapid modification of tactile signal features to match interaction scenarios -— for example, adding cascaded inflation to map to the needed quantity of a target ingredient in a recipe guidance task.

\subsection{Haptic Display Operational Modalities}
\label{sec:modalities}

In this work, the haptic displays operate using three distinct tactile feedback modalities: \textbf{pressure}, \textbf{frequency}, and \textbf{area of contact}. Each modality is enabled by a specific \textit{output logic}: pressure signals are controlled via individual addressing, frequency signals through the pneumatic oscillators, and area signals through cascaded inflation. While more complex modalities may be achievable by combining output logic stages, these tactile modalities represent those commonly seen in haptic displays. 

To characterize the behavior of each of these tactile feedback modalities, we conducted experiments measuring their dynamic response. For the \textbf{pressure} modality, the previously described valve characterization (Section~\ref{subsec:fluidic_logic}) shows that with an inlet pressure of $20.68$~kPa, the valves opens with a flow rate of $2.76$~slm with a rise time of $0.51$~seconds from the time the chamber starts being pressurized. 

For the \textbf{frequency} modality, oscillator characterization was conducted by connecting the \textit{oscillator} circuit to a haptic display and recording input/output pressures using pressure sensors (015PGAA5, Honeywell Sensing) and an outlet flow using a flow sensor (PFLOW3008, Angst+Pfister). The activation signal pressure $S$ was set to $20.68$~kPa, and the input pressure $P$ was increased from $0$~kPa until oscillations ceased ($75.84$~kPa) using a hand pressure regulator. Oscillations began at $P=22.41$~kPa, producing a frequency of $1.8$~Hz. A maximum stable frequency of $7.41$~Hz was reached at $P=75.84$~kPa, beyond which oscillations would slow and stop. Due to the rapid cycle time, outlet pressure and flow did not reach zero, resulting in pressure oscillations between approximately $3.48$ and $13.79$~kPa. 

For the \textbf{area of contact} modality, which relies on \textit{cascaded inflation} of valves, we can estimate the inflation delay using the earlier characterized valve flow rate. 
Assuming a valve chamber volume of $8.58$~mL at rest, an additional $1.75$~mL of air is needed to reach the snap-through pressure from atmospheric conditions. Adding the display pouch volume of $1.25$~mL (one of the pouches in the 3-DoF haptic display, estimated as $1/3$ of $3.75$~mL), the expected delay in cascading inflation would be $0.065$~seconds given the $2.76$~slm output flow of a valve. In practice, when measured using a flow sensor, delays were closer to $0.25$~seconds. The difference in between estimated and actual delay is likely due to pressure losses in the system and the time it takes to snap the first valve's chamber. As well, while we did not explicitly vary intermediate volumes to tune delay in this study, preliminary experiments suggest that increasing the volume of downstream elements (e.g., using a larger volume display for a wrist versus a finger) would lead to longer delays, an aspect which could be leveraged in future designs.



The control architecture described in Figure~\ref{fig:fluidic_logic} supports flexible reconfiguration of these output modalities through simple hardware swaps, without modifying the underlying software. While the configurations used in our user study were chosen for consistency between subjects, the system can scale to different code-to-output mappings by adjusting the logic inputs and signal blocks. Consider again the recipe guidance scenario: a haptic display for a user could signal the ingredient type with pressure, quantity with area, and correct placement with a high frequency buzz. Depending on the task demands or user preferences, the mapping can be reconfigured: more signals for ingredients can be added on the fly or the quantity signal could be removed completely for a more experienced user. Our modular fluidic logic system supports such personalized remapping without software changes.

A key advantage of our approach is the hardware modularity and ease of physical reconfiguration. All pneumatic connections between valves, displays, and pressure sources are made using standard Luer-lock barbed fittings, which enable quick and tool-free swapping of components. This plug-and-play design allows users to adapt the system layout relatively fast. 
This level of signal flexibility is further made possible by treating the output logic as modular signal blocks: if the soft valve assemblies show in Figure~\ref{fig:fluidic_logic} are pre-built, reconfiguring the haptic signals or swapping displays becomes a trivial task (see Supplemental Movie~2). While producing the soft valves requires some upfront effort, the long-term payoff is substantial: all logic components can be used in both code logic and output logic blocks and have shown little to no decline in performance over months of use. 



\section{Algorithmic Personalization} \label{sec:custom}

We have developed a modular haptic system that can be reconfigured to display a variety of signals. 
But how do we determine which configuration of the system is best for the current user? 
In this section we present the second piece of our modular approach: an algorithm that recommends mechanical deployments based on their potential to transfer information for a given user and task.
Our modular software approach considers both the task's perspective (i.e., the resolution of a signal) and the human's perspective (i.e., the user's preference for that signal).
We first formulate the problem setting, and then develop the information transfer approach.
We conclude this section with our human model, which enables users to provide their input on the types of signals they find most informative and interpretable.

\p{Problem Setting}
The device's objective is to convey task-relevant information $\theta \in \Theta$ to the human. 
Returning to our running example from \fig{front}, $\theta$ could be the next step in the recipe.
This information $\theta \sim \rho(\cdot)$ is sampled from the task's prior distribution $\rho(\theta)$, which encodes knowledge of the potential task-relevant information, i.e., the device can render an ingredient and a location but not two locations.
The device then displays a signal $s \in S$ (e.g., a change in pressure or a pulsating frequency). 
The human perceives this signal and attempts to complete the task based on their interpretation of the haptic feedback.
For clarity, let signal $s$ have $d$ dimensions, where each dimension corresponds to a different potential axis of variation for the signal (e.g., pressure, area, frequency). 

Our modular haptic device can be physically reconfigured to render a variety of different signals with different or the same dimensions.
More formally, let there be $N$ interface configurations $\mathcal{S} = \{{S}_1, \ldots, {S}_N\}$ we are interested in comparing.
Each configuration represents a different modular deployment that delivers its own range of haptic signals: $S_1$ may be a $1$-DoF display that renders pressure, and $S_2$ may be a $2$-DoF display that displays pressure and area.
Within our running example, when the fluidic logic is modified in \fig{front}, the haptic system in the new configuration $S_i$ now includes frequency in the signal dimension instead of pressure.
The purpose of our algorithmic personalization is to select the optimal configuration $S$ from the set of options $\mathcal{S}$.

\p{Information Transfer}
We emphasize that this optimal configuration is \textit{task}- and \textit{user}-dependent.
Some configurations may not have enough degrees of freedom to describe $\theta$, while another may be a better fit for the prior $\rho(\theta)$.
Additionally, certain users may find some signals salient and interpretable, while other users may find these same signals confusing, unintuitive, or otherwise distasteful \cite{nunez2019understanding, ikemoto2012physical}.
In order to determine the configuration that will be most effective for the current task and user, we turn to information theory \cite{shannon1948mathematical}. 
Information theory provides a principled way to assess how data is transferred; more specifically, we can use information theory to quantify how a given deployment of our modular haptic system will convey data to the human. 
The information-theoretically optimal configuration satisfies:
\begin{equation}
    S^\star = \argmax\limits_{S \in \mathcal{S}}
    I\left(\theta ~;~ s \mid S, \rho\right)
    \label{eq:mi}
\end{equation}
Here, $I$ is mutual information \cite{shannon1948mathematical}. Intuitively, \eq{mi} expresses how much information the human gains about $\theta$ after observing the signal $s$, given the current configuration $S$ and the task prior $\rho$. 
The configuration $S^\star$ that maximizes this mutual information will --- for any given data $\theta \sim \rho(\cdot)$ --- be most capable of selecting signals which convey the desired data to the human user. 

In what follows, we will formally show how \eq{mi} considers both task-specific details (i.e., the robot's perspective) and user-specific saliency (i.e., the human's perspective). By definition, mutual information in our setting can be separated into two components:
\begin{equation}
I\left(s ~;~ \theta \mid S, \rho\right)
= 
H\left(s \mid S, \rho\right) - H\left(s \mid \theta, S, \rho\right)
\label{eq:entropy}
\end{equation}
where $H$ is Shannon Entropy \cite{shannon1948mathematical}. 
To explain these components, we highlight that $H(x \mid y)$ captures the \textit{uncertainty} of variable $x$ given variable $y$. If $y$ fully captures $x$, then the entropy $H(x \mid y)$ is zero. On the other hand, if $y$ provides very little information about $x$, then $H(x \mid y)$ is high.

With this understanding in mind, the \textit{first term} of \eq{entropy} is absent of any user-specific personalization. Instead, it is a function of the probability of the signal occurring given the interface configuration and the task prior: 
\begin{equation}
H\left(s \mid S, \rho\right)
= 
- \sum p\left(s, S, \rho\right) \log p\left(s \mid S, \rho\right)
\end{equation}
Assuming that $\rho$ is independent of interface design, this reduces to a constant:
\begin{equation}
H\left(s \mid S, \rho\right) \propto \log\left(\lvert s \rvert \times \lvert S \rvert\right)
\label{eq:first-term}
\end{equation}
All else being held equal, an interface configuration with more signals that are highly descriptive (i.e., higher dimension) will have greater mutual information than a smaller interface with fewer signals. 
This is sensible: to convey data to the human, we should select deployments where the haptic system can display as much information as possible. 
Accordingly, \eq{first-term} captures the robot's side of information transfer --- selecting configurations that can render high-resolution signals.

The \textit{second term} of \eq{mi} takes the human's perspective into account, including aspects such as their preferences, their ability to parse signals, and their perception error:
\begin{equation}
H\left(s \mid \theta, S, \rho\right)
= 
- \sum p\left(s, \theta, S, \rho\right) \log p\left(s \mid \theta, S, \rho\right)
\label{eq:second-term}
\end{equation}
If the user decodes $\theta$ from $s$ without error, then \eq{second-term} is $0$. If they cannot determine $\theta$ based on the robot's signal $s$, then $H\left(s \mid \theta, S, \rho\right)$ approaches $\log \left(\lvert s \rvert \times \lvert S \rvert \right)$ and the mutual information from \eq{mi} becomes $0$. Intuitively put: if a given haptic display is able to render many different signals --- but the human cannot interpret any of them --- that configuration is useless in practice.
This mathematically highlights the importance of humans in information transfer; understanding how humans perceive different haptic signals is necessary to determine appropriate configurations.

\p{Human Preferences}
We encode the human's perception of a given haptic signal by modeling how people connect signals $s$ to data $\theta$.
Put another way, if the robot can convey a set of signals $S$, and the robot is trying to communicate the specific data $\theta$, what signal $s$ does the human expect the robot to display?
We write this relationship as the probability term $p(s \mid \theta, S, \rho)$ on the right side of \eq{second-term}.
To instantiate this probability, we consider user-specific saliency within the principle of maximum entropy \cite{jain2019probabilistic}:
\begin{equation}
    p(s \mid \theta, S, \rho) \propto 
    \exp\left(
    - \beta \cdot \|
    h(\theta) - s
    \|_W^2
    \right)
   \label{eq:human-model} 
\end{equation}
where $h:\Theta\mapsto S$ is a function representing the signal that the human expects for each $\theta$: i.e., this is how the human {thinks} they should interpret signals.
Within \eq{human-model} there are two parameters that specify the human's perception of the haptic signal.
First, constant $\beta \geq 0$ captures the overall \textit{sensitivity} of the user; second, matrix $W \in \mathbb{R}^{d \times d}$ encodes how \textit{preferable} the user finds each axis of the signal.
Putting each term together, \eq{human-model} says that humans find it easier to discern and interpret signals when they are sensitive to those signals (higher $\beta$) and when they prefer receiving those signals (larger $W$).
Of course, we recognize that this model is limited --- it does not explicitly account for affective or cognitive responses to feedback.
Future works can leverage our overarching modular framework and simply replace \eq{human-model} with other human models.
For our experiments, \eq{human-model} provides a way to quickly formulate user-specific preferences and represent the human's perspective of the haptic system.

By combining Equations~(\ref{eq:first-term}), (\ref{eq:second-term}) and (\ref{eq:human-model}) we ultimately reach information theoretic interfaces that balance \textit{utility} and \textit{interpretability}.
If a given configuration does not have a sufficient range of signals to convey the desired task (e.g., the display can only show two different signals, but there are five unique goals), then the first term in \eq{entropy} decreases and the information transfer falls.
Alternatively, if a given configuration renders signals that the human does not like or struggles to perceive (e.g., the haptic display can only apply different pressures, but the human cannot tell these pressures apart), then the second term in \eq{entropy} becomes more negative and again we fail to maximize information transfer.
The configuration $S^*$ that maximizes information transfer (a) provides sufficient signals to convey the task, while (b) aligning those signals with the human's preferences and perception.
We emphasize that the configuration that maximizes \eq{mi} is not simply the one that the user prefers.
As we show in our experiments, the optimal configuration must balance the user preferences with its inherent ability to convey data.

\p{Implementation}
\eq{mi} formally accomplishes this trade-off.
When applying \eq{mi}, we first record which signal types a given human prefers, and by how much, along a $\alpha$-$1$ scale, and then use their preferences to complete the matrix $W$.
Here higher values indicate increased preference.
Based on these results we can then compute the information gain for each possible haptic configuration $S \in \mathcal{S}$, and rank those configuration from most recommended to least recommended.
The configuration $S^*$ which maximizes \eq{mi} is then recommended to the user.
At this point the human can physically reconfigure the fluidic logic --- i.e., unplugging and plugging-in tubes --- to reach the recommended configuration and render signals that are both informative and intuitive.
\section{User Study}\label{sec:study}

\begin{figure*}
    \centering
    \includegraphics[width=2.0\columnwidth]{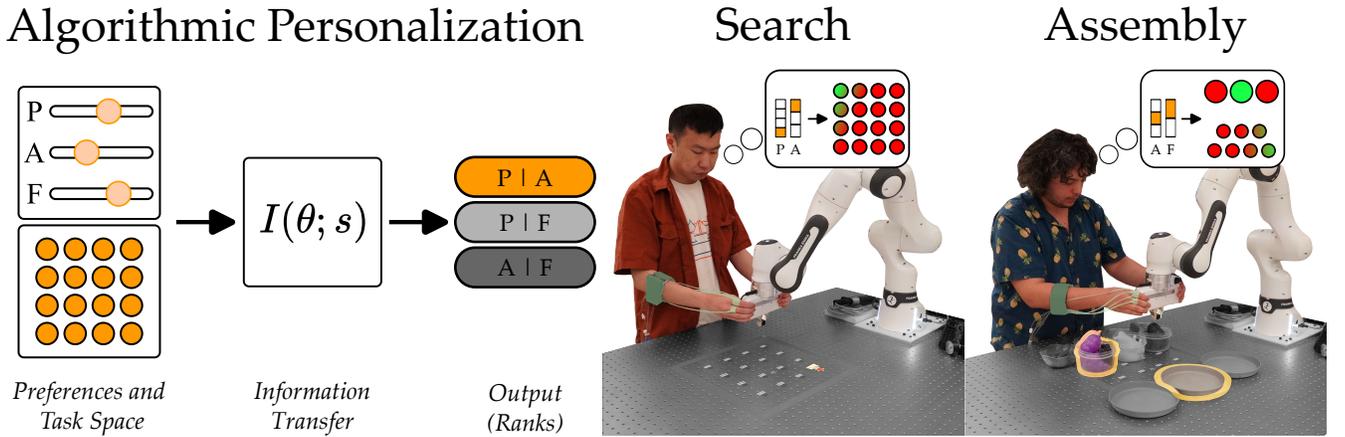}
    \caption{(Left) Our method for selecting ranking interface configurations. The algorithm ranks three interface configurations according to the user's preferences and task-specific parameters. This process is outlined in Section~\ref{sec:custom}.
    (Middle) A participant interacts with our modular haptic device in the \textit{Search} environment. 
    The haptic device is highlighted in green, and the task-relevant information is highlighted in orange. 
    (Right) A participant interacts with our modular haptic device in the \textit{Assembly} environment. The task-relevant information is highlighted in orange. 
    }
    \label{fig:exp_setup}
    \vspace{-0.5em}
\end{figure*}

\begin{figure}[t]
    \includegraphics[width=0.95\columnwidth]{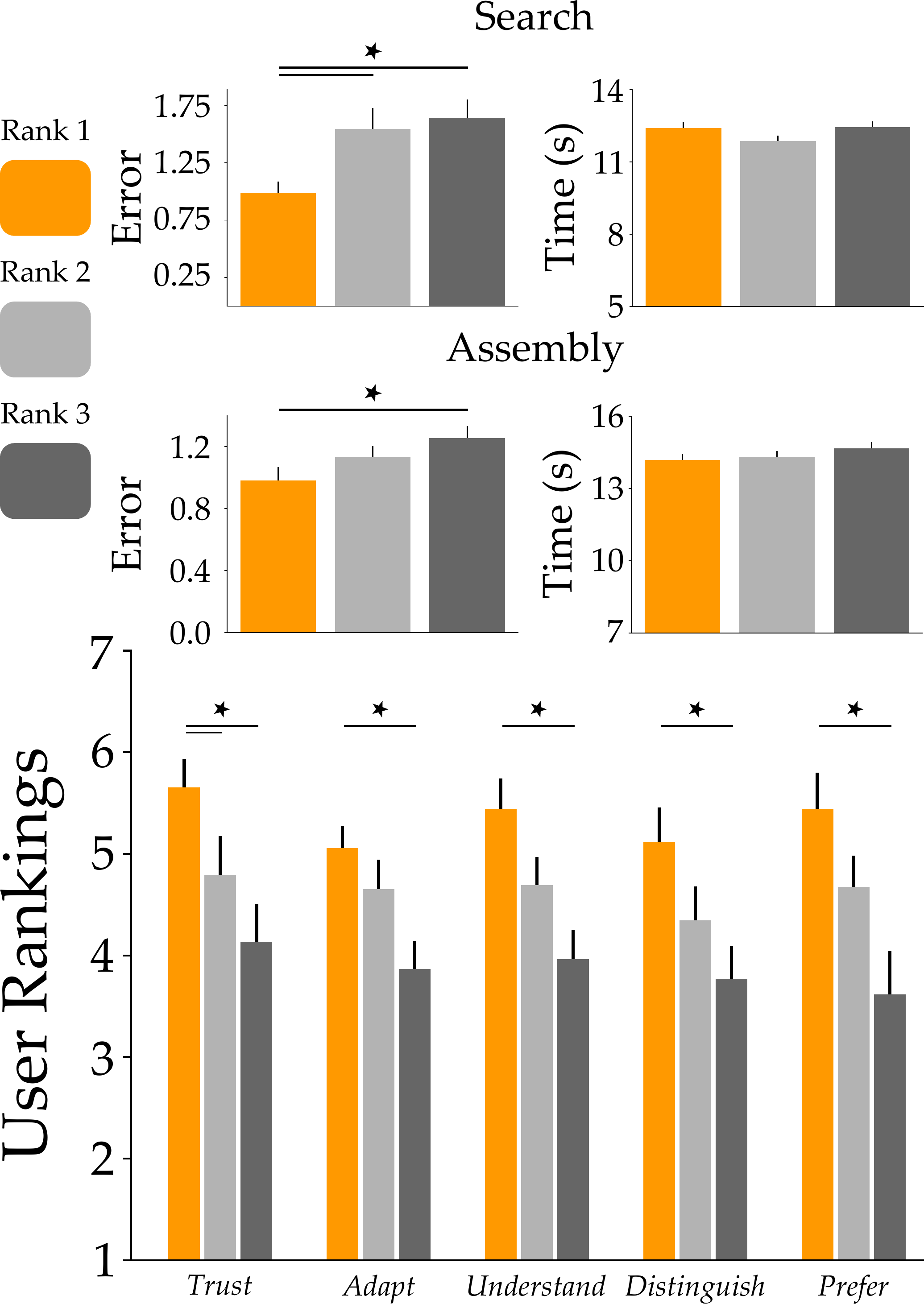}
    \centering
    \caption{Results from our in-person user study. An asterisk ($\star$) denotes significance. (Top) In the \textit{Search} task, 
    a repeated-measures ANOVA test revealed that configurations had a significant effect on error: $F(1.868, 192.454) = 6.719$, $p < .05$. Post-hoc analysis suggests that our predicted Rank~$1$ interface configuration outperformed the Rank~$2$ ($p < .05$) and Rank~$3$ configurations ($p < .001$) in terms of Error, but it does not outperform with respect to the Time needed to complete the task.
    (Middle) Likewise in the \textit{Assembly} task, a repeated-measures ANOVA test revealed that configurations had a significant effect on error: $F(1.984, 204.373) = 3.499$, $p < .05$. Post hoc-analysis shows that Rank $1$ configurations significantly outperform Rank~$3$ ($p < .05$) with a trend toward better performance over Rank~$2$ ($p = .197$).
    As in the \textit{Search} task, configurations did not seem to have a significant effect on time.
    (Bottom) Combined subjective ratings from our Likert-scale survey indicate a significant difference between the subjective preferences of Rank $1$, $2$, and $3$ configurations. A detailed analysis of results are presented in Table~\ref{table:likert}.
    }
    \vspace{-0.5em}
    \label{fig:results}
\end{figure}

\begin{figure}
    \centering
    \includegraphics[width=0.95\columnwidth]{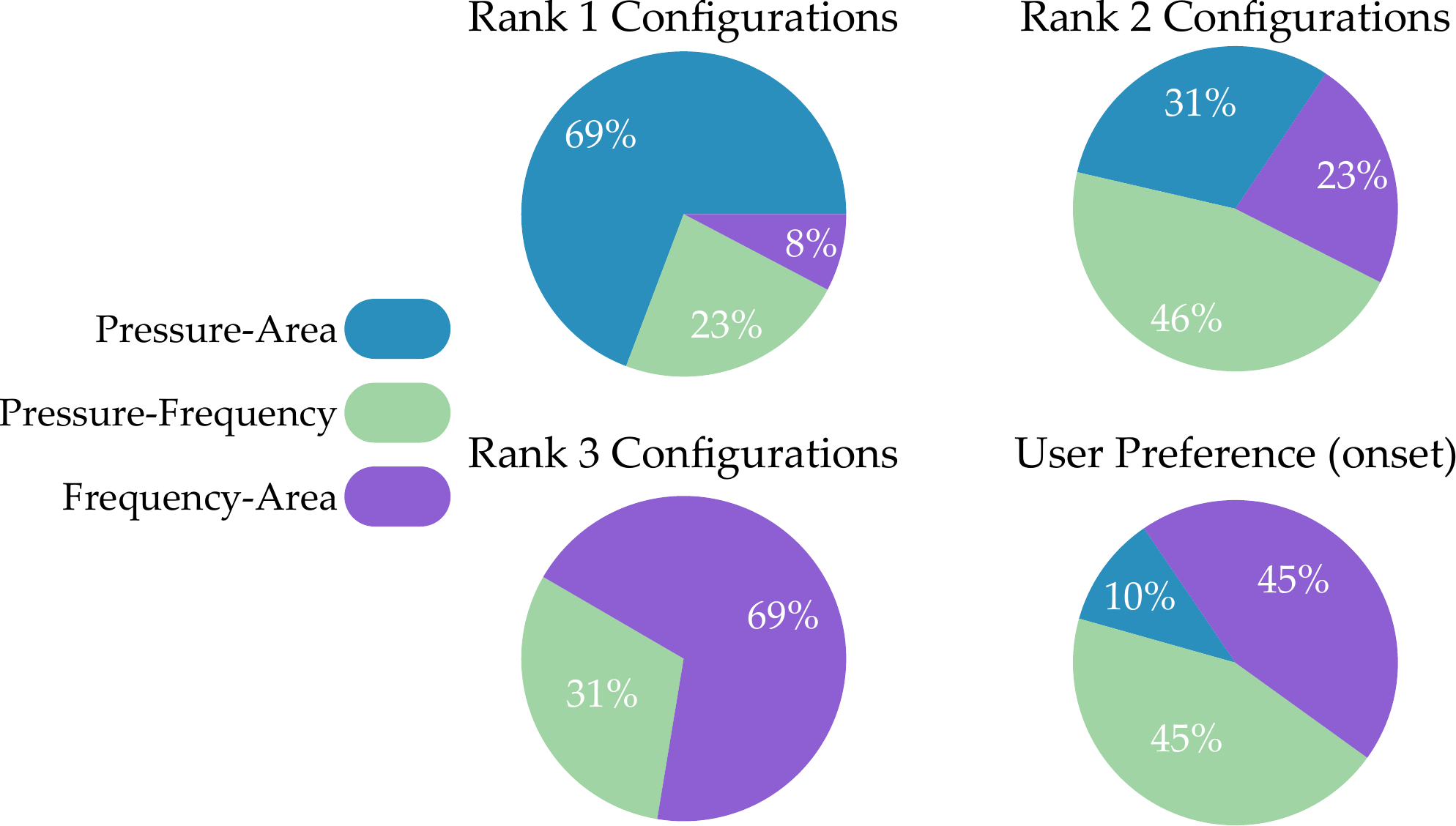}
    \caption{We report the number of times that each configuration is rated Rank $1$, $2$, or $3$. We also note the reported User Preference at the onset of the interaction. Note that this is \textit{not} the results of our Likert-Scale survey: 
    it is derived from the preference data we recorded prior to task completion. 
    These results show that although the algorithm in Section~\ref{sec:custom} accounts for preferences, it does not \textit{rely} on them: our method considers both user preference and task parameters. 
    If our method considered preferences alone, the User Preference and Rank $1$ Configuration charts would be identical. 
    Similarly, if preference data was ignored, then the Rank $1$ configuration chart would comprise of a single configuration.
    Instead, this figure shows that a diverse collection of interface configurations are deployed across users in a way that maximizes task performance while accommodating user preference.
    }
    \vspace{-0.5em}
    \label{fig:ranks}
\end{figure}

We evaluate our modular haptic system from Section~\ref{sec:device} and our personalization algorithm from Section~\ref{sec:custom} in a two-part user study with $n = 13$ members of the Purdue University community. In this study, we considered three modalities: pressure, area, and frequency (Section~\ref{sec:modalities}). To facilitate experimental design, enhance tractability, and support controlled comparison, we intentionally limited the number of discrete levels per modality. These constraints were also informed by practical considerations. Pressure signals were assigned four discrete levels ($6.89, 13.79, 20.68$ and $27.58$~kPa) due to the ease of implementing scalable valve configuration; area of contact signals use three discrete levels, reflecting the maximum number of independently controlled pouches in our 3-DoF display; and frequency signals were limited to two levels ($4$ and $7$~Hz), based on the range of stable frequencies of the pneumatic oscillators. In each of two tasks, the user interacts with three combinations of these modalities: pressure-area (PA), pressure-frequency (PF), and area-frequency (AF).

All three configurations used a distinct combination of code logic and output logic circuits. For each configuration, the user interacted with two separate ring displays --- one for each modality in the pair. The pressure and frequency modalities used the 1-DoF displays, while the area modality used the 3-DoF display. This setup allowed us to isolate the effect of each modality pairing while maintaining system consistency and modular integrity. 

\p{Experimental Setup}
Participants performed two tasks involving physically guiding a Franka Research 3 manipulator with the feedback provided by our modular haptic device. Before beginning, we showed all the modalities and their respective levels to users. Participants indicated their modality preferences using a sliding scale interface (see \fig{exp_setup}). 
These preferences were then converted into a preference matrix according to \eq{saliency} for use in our human model (\eq{human-model}):
\begin{equation}
    W(\alpha) = \frac{1}{1 + \alpha} 
    \begin{bmatrix}
    P_1 + \alpha & 0 & 0 \\ 
    0 & P_2 + \alpha & 0 \\ 
    0 & 0 & P_3 + \alpha
    \end{bmatrix}
    \label{eq:saliency}
\end{equation}
\eq{saliency} normalizes saliency in the range $[\alpha, 1]$. We chose $\alpha = .25$ for this user study. 
Using the personalization algorithm detailed in Section~\ref{sec:custom}, we ranked the three interface configurations for each user based on their expected information gain. These configurations consisted of two haptic displays, each conveying signals from one modality mapped to an axis of information in the task. We identified these configurations per user as Rank~$1$, $2$, and $3$, corresponding to the highest, intermediate, and lowest expected information gains, respectively.
In both tasks, the robot manipulator has hidden information that it communicated through the modular haptic system using these configurations. 
The first task, \textit{Search}, was a goal-reaching task: the robotic manipulator had a discrete target position within a $4\times4$ grid. Users interpreted haptic signals to guide the manipulator's end effector toward the perceived intended goal. 
The second task, \textit{Assembly}, was a higher-dimensional ingredient-selection scenario. Here, the hidden information specified one of seven ingredients placed in individual bowls and one of three destination plates. Participants interpreted the haptic cues, moved the robot's end effector to select the correct ingredient bowl, and subsequently placed the ingredient onto the indicated plate.

\p{Participants and Procedure}
We recruited $13$ participants ($6$ female, average age $24 \pm 4$) of the Purdue University community for this user study. Of the $13$ participants, five had not previously used robotic manipulators and $10$ had not previously interacted with haptic devices. Participants provided written consent according to university guidelines (IRB \# 2025-19).

We leveraged a within-subjects design, with each participant completing $8$ trials per interface configuration. Participants were not informed about the relative ranking or expected performance of each interface configuration (e.g., which interface had higher information gain). To mitigate order effects, we randomized and counterbalanced both the sequence of tasks and the order of interface configurations across participants. For instance, approximately half of the participants began with the \textit{Search} task, and about a third started with the pressure-area configuration. 

In each trial, participants received two consecutive haptic signals, each lasting three seconds, with each signal corresponding to a distinct piece of task-relevant information. Specifically, in the \textit{Search} task, the first signal corresponded to the target’s discrete position along the X-axis, and the second to the position along the Y-axis. Similarly, in the \textit{Assembly} task, the initial signal indicated which ingredient participants should select from among seven options, while the subsequent signal indicated the plate onto which the ingredient should be placed from three possible locations. Participants were instructed that signal intensity corresponded proportionally to magnitude; for example, greater signal intensity indicated a more distant target location along the respective axis. After each trial, participants received feedback indicating the accuracy of their responses, including the correct target and the Manhattan distance between their selection and the intended answer. The entire user study procedure lasted approximately $50$ minutes per participant.

\p{Dependent Measures --- Objective}
We leveraged the Franka Research 3 manipulator to measure the position of the end-effector. To assess user performance, we considered two metrics: \textit{time} and \textit{error}. 
In the \textit{Search} task, error was quantified as the mean-squared error (MSE) between the intended goal position $\theta$ and the center of the closest grid cell to the user's final end-effector position. 
In the \textit{Assembly} task, we calculate error in a similar way by determining the closest bowl and plate during the interaction and measuring the mean-squared error of the intended bowl and plate with the recorded bowl-plate combination.

\p{Dependent Measures --- Subjective}
Between each configuration, participants completed a 7-point Likert scale survey. This survey assessed the user's subjective preferences along five multi-item scales. We asked users:
\begin{enumerate}
    \item If they \textit{trusted} the interface to guide them.
    \item If they \textit{adapted} to the interface's signals.
    \item If they \textit{understood} the interface's signals.
    \item If they could \textit{differentiate} the interface's signals.
    \item If they \textit{preferred} using this configuration.
\end{enumerate}

\begin{table*}[t]
\caption{Questions from our Likert-scale survey grouped into five scales: Trust, Adaptation, Intuitiveness, Clarity, and Preference. The reliability (Cronbach's $\alpha$) and results from a repeated-measures ANOVA are reported for each scale. A $p<.05$ indicates significant differences across methods.}
\label{table:likert}
\centering
\begin{tabular}{lccc}
\hline
Questionnaire Items & Reliability & F($2$,$50$) & p-value \\
\hline
Trust & $.87$ & $8.7$5 & $<.001$ \\
\quad I trusted the interface to guide me accurately during the task. & & & \\
\quad I could not rely on the interface's signals when completing this task. & & & \\
\hline
Adapt & $.69$ & $7.15$ & $<.01$ \\
\quad I was able to adapt to the feedback provided by this configuration. & & & \\
\quad I struggled to adapt to the interface's signals. & & & \\
\hline
Understand & $.76$ & $9.59$ & $<.001$ \\
\quad I understood what this interface was trying to convey. & & & \\
\quad I found the signals from this configuration confusing and unclear. & & & \\
\hline
Distinguish & $.85$ & $5.03$ & $<.05$ \\
\quad I could easily differentiate between signals when using this interface. & & & \\
\quad I could not distinguish between signals when using this configuration. & & & \\
\hline
Prefer & $.91$ & $10.18$ & $<.001$ \\
\quad I would prefer to use this configuration again in future tasks. & & & \\
\quad I would prefer to not use these modalities for future tasks. & & & \\
\hline
\end{tabular}
\vspace{-0.5em}
\end{table*}

\p{Hypotheses}
We had three hypotheses for this user study:
\begin{quote}
    \textbf{H1.} The configurations that our method selects will be more \textit{efficient} than alternatives. 
\end{quote}
\begin{quote}
    \textbf{H2.} The configurations that our method selects will be more \textit{understandable} than alternatives. 
\end{quote}
\begin{quote}
    \textbf{H3.} Users will \textit{subjectively prefer} configurations that our method selects. 
\end{quote}

\p{Results -- Objective}
We analyzed user performance in both the \textit{Search} and \textit{Assembly} tasks using two objective measures: error and completion time. We conducted repeated-measures ANOVA tests with configuration (Ranks~$1$–$3$) as the within-subjects factor. We report significant effects at $p < .05$ and apply Bonferroni correction for post-hoc comparisons.

In the \textit{Search} task, we found that the configuration of the haptic display had a significant effect on error ($F(1.868,\ 192.454) = 6.719$, $p < .01$). Post-hoc comparisons confirmed that Rank~$1$ (the configuration recommended by our method) significantly outperformed Rank~$2$ ($p < 0.05$) and Rank~3 ($p < .001$). Completion time did not differ significantly across configurations ($F(2,\ 206) = 2.518$, $p = .083$), with Rank~$2$ yielding the shortest average time.

In the \textit{Assembly} task we again observed a significant effect of configuration on error ($F(1.984,\ 204.373) = 3.499$, $p < .05$). Post-hoc analysis showed that Rank~$1$ significantly outperformed Rank~$3$ ($p < .05$), with Rank~$1$ showing a non-significant trend toward better performance compared to Rank~$2$ ($p = .10$). Completion times did not show a significant main effect ($F(2,\ 206) = 2.369$, $p = .096$).  While these results indicate no statistical difference in time across configurations, this can be viewed positively: improved task accuracy with Rank~$1$ configurations was not achieved at the cost of longer task durations. These results suggest that the gains in accuracy introduced by the personalized Rank~$1$ configurations were not due to slower, more deliberate actions, but instead reflect improved signal clarity and interpretability. 

\p{Results -- Subjective}
After using each haptic configuration, participants rated their experience across five scales: Trust, Adaptation, Intuitiveness, Clarity, and Preference. Table~(\ref{table:likert}) and Figures~(\ref{fig:results}-\ref{fig:ranks}) summarize our results.
We first tested the reliability of our five scales and found that all five were reliable (Cronbach's $\alpha > 0.7$). Accordingly, we grouped each of these scales into a combined scale and performed a repeated-measures ANOVA on the results. 
The haptic configurations recommended by our modular approach consistently received the highest ratings across all subjective measures ($p < 0.05$). Participants reported greater trust in Rank~$1$, found it easier to adapt to, and described it as clearer and more intuitive. Preference scores also strongly favored Rank~$1$, indicating a high level of user satisfaction with the configurations selected by our algorithm. These results suggest that users could not only perform better with Rank~$1$, but also subjectively recognize and value its benefits. 

\section{Discussion}
\label{sec:discussion}
The user study results demonstrate that both aspects of our modular approach led to improved objective performance and more favorable user experiences. Rank~$1$ configurations (i.e., hardware configurations recommended by our information gain algorithm) consistently resulted in lower task error and higher subjective ratings, supporting our hypotheses.

\p{Impact of Hardware Modularity} 
Our modular hardware approach enables physical reconfiguration of the haptic displays, which proves essential in accommodating individual differences. While in the user study the PA configuration was most frequently selected as Rank~$1$ by our personalization algorithm, the other displays were better for $4$ out of $13$ users. Without hardware modularity, this result would suggest that PA is the most effective interface out of the three options to build. However, with the modular and reconfigurable hardware, each participant could be given their Rank~$1$ interface configuration. 
While not exercised to the fullest extent in this experiment, this reinforces the value of physical reconfiguration: a larger variety of different combinations of pressure, frequency, and area could feasibly allow the system to further accommodate diverse user needs, preferences, and perceptual sensitivities with only quick hardware reconfigurations. Without this mechanical flexibility, such personalization would not be possible. 


Additionally, modality effectiveness did not always fully transfer across tasks. In the \textit{Assembly} task --- which required interpreting one signal to choose an object from a set of seven and another from a set of three --- participants found certain configurations less intuitive. For instance, one noted, ``area was super hard when it was assigned to finding the ingredient,'' and another stated, ``the assembly task was difficult and frustrating because none of the modalities mapped well.'' These experiences emphasize that signal-task alignment is critical as well. Here, hardware modularity offers another key advantage: it allows designers to quickly reconfigure which modalities are assigned to which task components (e.g., mapping a clearer signal to the harder selection task) and the dimensionality of those modalities.

Beyond supporting personalization in this study, the hardware modularity of our system plays a critical role in enabling broader adaptability. From a practical standpoint, swapping in new signals types or adjusting modality pairing requires minimal effort --- standardized valve assemblies and interchangeable displays could allow researchers to rapidly prototype and evaluate new configurations iteratively, without modifying software or rebuilding the core system. This potentially makes our system scalable to more complex interfaces and to user adaptation over time. 

\p{Impact of Algorithmic Recommendations}

Complementing the mechanical flexibility, our algorithmic layer determined which hardware configuration to deploy for each user. Interestingly, the configurations selected by the algorithm did not always match participants' initial preferences, nor pick the display with the most distinct signals. Figure~\ref{fig:ranks} illustrates this: if our algorithm relied only on preferences, the Rank~1 and User Preference distributions would be identical --- but they are not. If it ignored preferences entirely, all Rank~$1$ configurations would have been the same --- which is also not the case. Instead, the algorithm successfully considers both user preferences and task parameters, selecting the signal pair that is expected to maximize information transfer for each user. As a result, PA appeared as the top-ranked interface for 9 out of $13$ participants and the remaining participants were assigned either PF ($3$ participants) or AF ($1$ participant) as their top configuration. In all cases, each participant --- regardless of their assigned configuration --- performed best with their Rank~$1$ interface configuration, keeping the same trends seen in \fig{results}.

Some reasons for the discrepancy from user preference or maximal signal are suggested by participant comments. Some participants reported perceptual overlap between modalities, making certain configurations harder to interpret. For example, one user remarked, ``PA confused me a little because they felt like very similar signals,'' while another shared, ``I thought I had pressure figured out but then I would make mistakes that would make me doubt it.'' 
Other participant feedback highlighted where saliency was high but granularity was low. One participant wished that ``frequency had more levels,'' while another offered a more strategic suggestion: ``I would like to separate the ingredients into two groups, then I would assign two modalities to the ingredients only.'' 

It is important to note that this is with a simple static human model, i.e. user reported preference after a single interaction, and with a small set of task repetitions without time for significant adaptation. 
Interestingly, one participant described their experience of decoding haptic signals by drawing parallels to the process of machine learning. ``So this is how it feels to be a machine learning algorithm?'' This reaction highlights a core insight of our approach: participants were engaging in a form of online inference, forming and refining internal models of the interface over time. This perceptual co-adaptation suggests that future systems could go beyond static personalization, incorporating metrics like user error and reaction time to continuously re-rank or adjust signal selection over time.

\p{System Simplicity and Limitations}
The pneumatic infrastructure required to support our reconfigurable haptic system is fundamentally simple: a pressurized air supply, hand pressure regulators to produce $P_0$ to $P_N$ and signal level pressure, solenoids to turn on and off code logic inputs, and something to control the on/off state of the solenoids, such as a microcontroller or logic switches. These relatively inexpensive components and simple control makes the system easily accessible and maintainable for users without previous experience in pneumatic device design. That said, there are limitations when trying to scale to higher input-output combinations. Although our $3I/8O$ system performed reliably, increasing complexity to, for example, $5I/32O$ or beyond may begin to introduce significant mechanical delays as valves respond sequentially, and pressure losses in longer chains may impact consistency. Higher efficiency and faster switching mechanical valves could aid in such scaling \cite{stanley2024high}. 
Another potential area for improvement within our modified soft valve design would be an option to maintain full bistability, as seen in the original molded design. 
In our version, the use of stock silicone tubing introduces greater stiffness to the internal structure, causing the membrane to be monostable when depressurized. For many logic operations in our system, this behavior is sufficient and even desirable. However, for more complex display behaviors could be achieved with persistent state memory and asynchronous behavior, which a fully bistable valve would allow. 


\section{Conclusion} \label{sec:conclusion}

In this work we presented a modular haptic display with reconfigurable signals.
The basis of our haptic display was a set of fluidic logic circuits.
By reconfiguring, i.e. connecting and disconnecting, these logic circuits, users can manually change the haptic signals that their interface renders.
A key implementation benefit was that --- because of the design of these circuits --- no pressure or flow controllers were required, and, in our user experiments, the same software could provide combinations of either pressure signals, frequency signals, or area signals with only hardware reconfiguration.
Overall, our physically modular hardware set the stage for haptic systems that can customize their signals to align with the current user and task.
In practice, this modularity enabled a wide range of different configurations; to determine which configuration was best for the current user and task, we formulated a set of equations to recommend the haptic configuration that maximized information transfer.
This recommendation balanced both the robot's perspective (choosing information-rich displays) and the human's perspective (selecting displays that the human could sense and prefer).
Our experiments across two tasks suggest that haptic signals which were personalized to the user led to more effective performance than the alternatives.
Both aspects of our work contributed to its overall effectiveness: (a) different individual users performed better with different configurations, confirming that one size does not fit all and mechanical customization is required, and (b) the recommended --- and best performing --- interface is not simply the haptic display that user's subjectively prefer. 
Instead, our algorithm correctly balanced user preferences with the task's communication requirements.
Overall, our mechanical, algorithmic, and experimental findings suggested that personalizing haptic systems through modular, reconfigurable signals enables more seamless human-machine communication.

\appendix 
\renewcommand\thefigure{A\arabic{figure}}
\setcounter{figure}{0}
\subsection{Code Logic Details}
\label{appendix}

\begin{figure}[h]
\centering
\includegraphics[width=0.8\columnwidth]{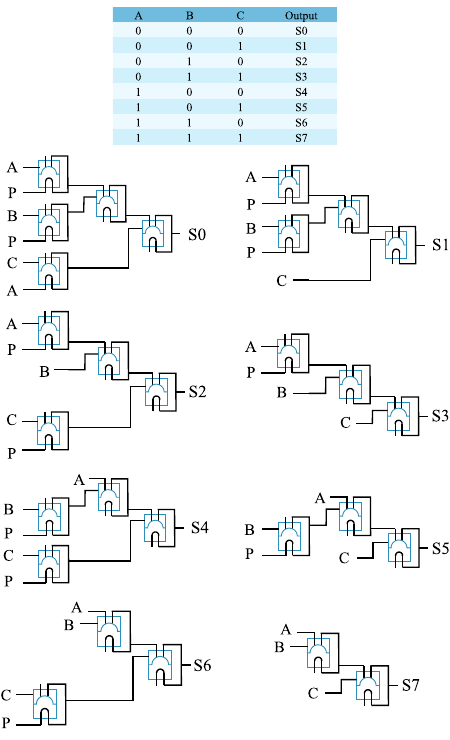}
\vspace{-0.5em}
\caption{Fluidic logic circuits automatically generated using the Soft Compiler developed by Kendre et al.~\cite{kendre2022soft}. These circuits implement the code logic stage of our fluidic control architecture, functioning as a pneumatic demultiplexer. A 3-bit input (A, B, C) selects one of eight output states (S0–S7), which in turn activates a downstream valve that routes actuation pressure ($P_0, P_1, ..., P_N$) to the output logic stage.}
\label{fig:fluidic_logic_2}
\vspace{-0.5em}
\end{figure}

Here we provide additional details for fluidic logic circuits implemented in the code logic. Figure~\ref{fig:fluidic_logic_2} shows the set of fluidic logic circuits used to implement our 3-bit pneumatic code logic. These circuits were automatically generated using the Soft Compiler tool developed by Kendre et al.~\cite{kendre2022soft}, based on the truth table shown. Each output state (S0–S7) corresponds to one unique combination of inputs (A, B, C), and the resulting signal is used to activate a separate valve that passes actuation pressure to the output logic stage, as shown in Figure~\ref{fig:fluidic_logic}.

\bibliographystyle{IEEEtran}
\bibliography{references}


\vspace{-1.5cm}

\end{document}